# Celestial Mechanics and Polarization Optics of the Kordylewski Dust Cloud in the Earth-Moon Lagrange Point L5

# Part II. Imaging Polarimetric Observation: New Evidence for the Existence of Kordylewski Dust Cloud

Judit Slíz-Balogh[1,2], András Barta[2,3] and Gábor Horváth[2,*]

1: Department of Astronomy, ELTE Eötvös Loránd University,
H-1117 Budapest, Pázmány sétány 1, Hungary

2: Environmental Optics Laboratory, Department of Biological Physics, ELTE Eötvös Loránd University, H-1117 Budapest, Pázmány sétány 1, Hungary

3: Estrato Research and Development Ltd., H-1124 Némétvölgyi út 91/c, Budapest, Hungary

*: Corresponding author, e-mail address: gh@arago.elte.hu

## Abstract

Telescopes mounted with polarizers can study the neutral points of the Earths atmosphere, the solar corona, the surface of planets/moons of the Solar System, distant stars, galaxies and nebulae. These examples demonstrate well that polarimetry is a useful technique to gather astronomical information from spatially extended phenomena. There are two enigmatic celestial objects that can also effectively be studied with imaging polarimetry, namely the Kordylewski dust clouds (KDCs) positioned around the L4 and L5 triangular Lagrangian libration points of the Earth-Moon system. Although in 1961 the Polish astronomer, Kazimierz Kordylewski had observed two bright patches near the L5 point with photography, many astronomers assume that these dust clouds do not exist, because the gravitational perturbation of the Sun, solar wind and other planets may disrupt the stabilizing effect of the L4 and L5 Lagrange points of the Earth and Moon. Using ground-born imaging polarimetry, we present here new observational evidence for the existence of the KDC around the L5 point of the Earth-Moon system. Excluding artefacts induced by the telescope, cirrus clouds or condensation trails of airplanes, the only explanation remains the polarized scattering of sunlight on the particles collected around the L5 point. By our polarimetric detection of the KDC we think it is appropriate to reconsider the pioneering photometric observation of Kordylewski. Our polarimetric evidence is supported by the results of simulation of dust cloud formation in the L5 point of the Earth-Moon system presented in the first part of this paper.

**Keywords:** celestial mechanics, polarization, Earth, Moon, instrumentation: polarimeters, methods: observational

# 1. Introduction

In astronomy, the majority of knowledge originates from information obtained via light. Although light is a transversally polarized electromagnetic wave (Azzam & Bashara 1992), astronomical information is collected mainly with telescopes detecting only the spectrum (radiance and colour) of the light of celestial objects within a limited wavelength range without polarization. Due to the polarization insensitivity of the majority of telescope detectors, valuable astronomical information remains unrevealed/undetected.

Fortunately, a few telescopes are mounted with linear and/or circular polarizers and can also measure the polarization characteristics of light of distant celestial objects, not just their spectrum. The nearest celestial phenomenon of semi-astronomical importance is the unpolarized (polarizationally neutral) points of the Earth's atmosphere, namely the Arago's, Babinet's, Brewster's and the fourth neutral points observed first in 1809 (Arago 1811), 1840 (Babinet 1840), 1842 (Brewster 1842, 1847) and 2001 (Horváth et al. 2002). Nowadays these celestial points are studied with imaging polarimetry, a very useful technique to gather information from spatially extended phenomena in the optical environment (Horváth & Varjú 2004; Horváth 2014). Farther targets of astronomical imaging polarimetry are the Sun, its planets and their moons in the Solar System (Gehrels 1974; Können 1985; Belskaya et al. 2012). Although the direct sunlight is unpolarized, the solar corona is partially polarized due to Compton scattering on the electrons of the Sun's atmosphere (Können 1985). The polarization pattern of the solar corona can be measured, if the bright Sun's disc is artificially occluded by an opaque disc, or when the Moon occludes it during total solar eclipses (Können 1985; Horváth & Varjú 2004). Planets and moons reflect partially polarized light, from the polarization characteristics of which certain surface features can be revealed that would be hidden for polarization-blind telescopes. Much farther targets are various stars, comets, galaxies and nebulae, the light of which is originally more or less polarized or becomes polarized due to interstellar and intergalactic magnetic fields (Gehrels 1974; Marin et al. 2012; Hadamcik et al. 2014; Reig et al. 2014; Ivanova et al. 2015; Marin et al. 2015; Zejmo et al. 2017).

These examples demonstrate well the usefulness of polarimetry in astronomy, the targets of which range from the nearest neutral points of the Earth's atmosphere up to the farthest nebulae. However, farther than these neutral points, but nearer than the Sun, there are two enigmatic celestial objects that can also effectively be studied with imaging polarimetry, namely the Kordylewski dust clouds (KDCs) positioned around the L4 and L5 triangular Lagrangian libration points (called simply Lagrange points further on) of the Earth-Moon system (Steg & De Vries 1966; Szebehely 1967; Danby 1992). In spite of the fact that in 1961 the Polish astronomer, Kazimierz Kordylewski had observed the dust cloud near the L5 point with photography (Kordylewski 1961) and photographs have been taken from the western hemisphere of the Earth confirming the existence of the KDCs (Simpson 1967), many astronomers do not believe this (Roosen et al. 1967; Bruman 1969; Valdes & Freitas 1983; Igenbergs et al. 2012). The skeptics assume the KDCs do not exist, because the gravitational perturbation of the Sun, solar wind and other planets has too strong a destabilizing effect on the L4 and L5 Lagrange points of the Earth and Moon. Thus, the existence together with photometric documentation of the KDCs is considered with disbelief by many astronomers.

Using ground-born imaging polarimetry, in this work we present the polarization patterns of the KDC observed around the L5 Lagrange point of the Earth-Moon system. We show that these patterns cannot be explained by any other optical phenomena, such as reflection artefacts within the telescope, cirrus clouds or condensation trails of airplanes. Excluding all possible artefacts, the only convincing explanation remains the polarized scattering of direct sunlight on dust particles collected in the L5 point. By our polarimetric observation we present here new optical evidence of the existence of the KDC, and by this we think it is appropriate to reconsider the pioneering photometric observation of Kordylewski (1961). Our polarimetric evidence is supported by the results of computer simulation of dust cloud formation aroundthe L5 point of the Earth-Moon system presented in the first part (Slíz-Balogh et al. 2018) of this paper.

## 2. Materials and Methods

As the KDCs are a very faint phenomenon, instead of simple photometry, we applied sequential imaging polarimetry, in which three necessary polarization pictures are taken one after the other. Our imaging polarimetric measurements have been performed in the private observatory of one of the authors (SBJ) located in Badacsonytördemic (east longitude $17^o\ 28'\ 15"$, north latitude $46^o\ 48'\ 27"$), Hungary. We used a Tokina AF 300/2.8 teleobjective equipped with a Moravian G3-11000 ABG full-frame CCD camera with a field of view of $7.5^o$ (horizontal) × $5^o$ (vertical) with a filter wheel for three 2 inch linearly polarizing filters (Edmund Optics, 43-785, USA), the transmission axis of which was rotated by $120^o$ relative to each other. At the chosen celestial region, using this imaging polarimetric telescope we took one photograph through each polarizing filter with an exposure of 180 seconds. In this work all points of time will be given as the mean time (middle of the exposure) of the second polarization picture. These three polarization pictures were evaluated with the use of the software AlgoNet (http://www.estrato.hu/algonet). The software computed the radiance $I$ as well as the degree $p$ and angle $\alpha$ of linear polarization in every pixel of the picture in the red (650 nm), green (550 nm) and blue (450 nm) part of the visible spectrum. The evaluated optical data $I$, $p$ and $\alpha$ are visualized with colour-coded two-dimensional distribution sky-maps/patterns in the three colour channels. Further details of imaging polarimetry and evaluation of polarization pictures can be found elsewhere (Horváth & Varjú 2004).

We took polarization pictures of the L5 point of the Earth-Moon system and its small ($7.5^o \times 5^o$) surrounding under the following different conditions: (i) The sky was cloudless. (ii) The studied small sky region was covered by a thin cirrus cloud. (iii) Condensation trails of an airplane occurred in this celestial window. (iv) As an important control, we photographed the same celestial region in the Taurus constellation through the three polarizers when the L5 point was not in this window.

To minimize the noise, the camera was cooled to -10 $^o$C and a dark frame (complementing with bias frame) was used. The dark frame was subtracted from the image to minimize the effect of noise, while the rest of the image was unaffected. To avoid the brightness variation due to vignetting, we used a flat-field correction. For image processing we used MaxIm DL6. Onto the motor-driven German Equatorial Fornax 100 Telescope Mount an extremely accurate driving correction system called Telescope Drive Master (TDM) was installed.

## 3. Results

After several-months of perseverance (because it is hard to find moonless and cloudless good nights in Hungary) we succeeded in catching the KDC around the L5 Lagrange point on two consecutive nights. Several photographs were takenon both nights, some of them were suspected of having thin clouds or vapour in the observation window. The latter measurements were rejected. Figure 1 shows the positions of the Moon and the L5 Lagrange point of the Earth-Moon system in the plane of the Moon's orbit when our imaging polarimetric measurements were taken.

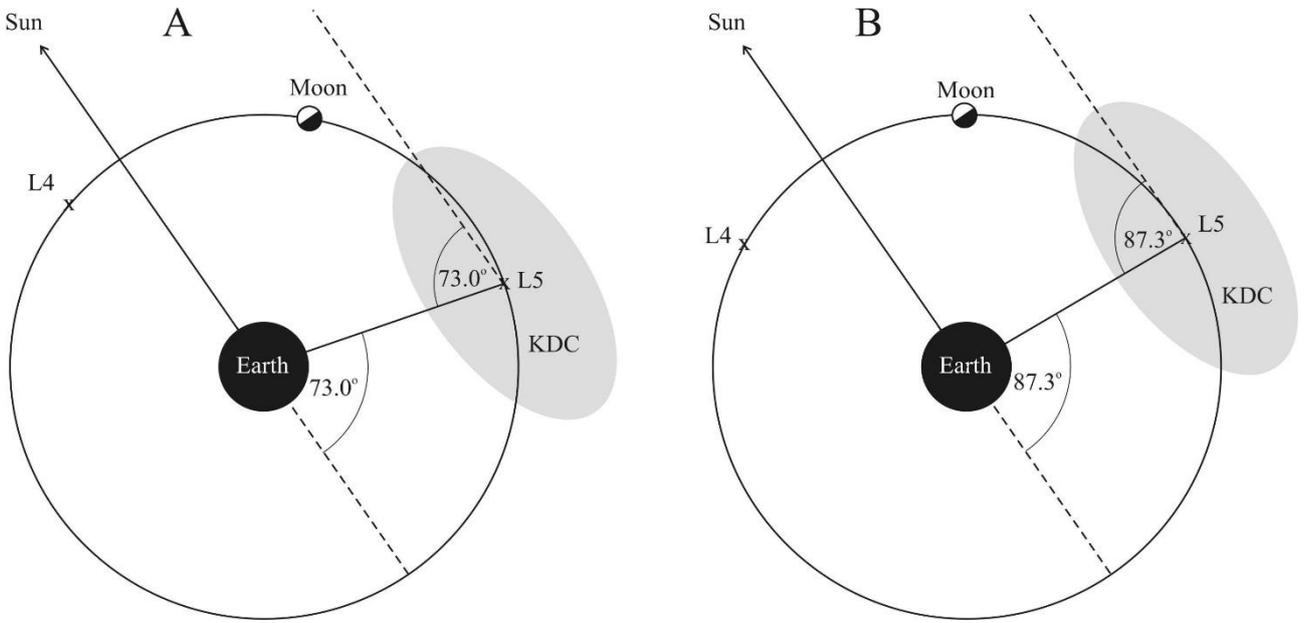

**Figure 1**: Positions of the Moon and the L5 Lagrange point of the Earth-Moon system in the plane of the Moon's orbit on 17 August 2017 at 23:29:67 UT with 73.0° phase angle (A), and on 19 August 2017 at 01:14:15 UT with 87.3° phase angle (B). Apart from the Earth and Moon, the relative dimensions are not to scale. The Sun's direction is indicated by an arrow. KDC: Kordylewski dust cloud.

Figure 2(A) displays the optical information gathered about the KDC with imaging polarimetry in the green (550 nm) spectral range at 23:29:67 UT on 17 August 2017. In the colour photograph and the pattern of radiance $I$, only a gradual brightening toward the Sun is visible due to the sunlight scattered by the dust particles. In the pattern of the degree of polarization $p$, two diffuse clusters of black pixels with $10\% \leq p \leq 20\%$ are discernible: one around the L5 point and another near the top right corner of the pattern. Both clusters of red pixels with $81° \leq \alpha \leq 90°$ are also seen in the pattern of the angle of polarization $\alpha$. In the red (650 nm) and blue (450 nm) spectral ranges the $I$-, $p$- and $\alpha$-patterns were very similar to those measured in the green part of the spectrum (Figure 2(A)). It is a very important feature that all local directions of polarization (short white bars) are perpendicular to the plane of scattering. This expected characteristic is typical for all scattering phenomena and proves that around the L5 point many light-scattering centers exist. These scatterers cannot be anything other than dust particles. Hence, in Figure 2(A) the KDC has a bipartite innate structure.

Figure 2(B) shows the $I$-, $p$- and $\alpha$-patterns of the KDC around the L5 point as detected in the green (550 nm) spectral range by imaging polarimetry at 01:14:15 UT on 19 August 2017. The characteristics of the dust cloud are similar to those in Figure 2(A) apart from the structure composed of five smaller diffuse clusters with $10\% \leq p \leq 20\%$ and $81° \leq \alpha \leq 90°$. This demonstrates that the structure of the dust cloud may change temporally. Thus, the KDC might be a dynamic structure as expected on the basis of computer simulations presented in the first part of this paper (Slíz-Balogh et al. 2018).

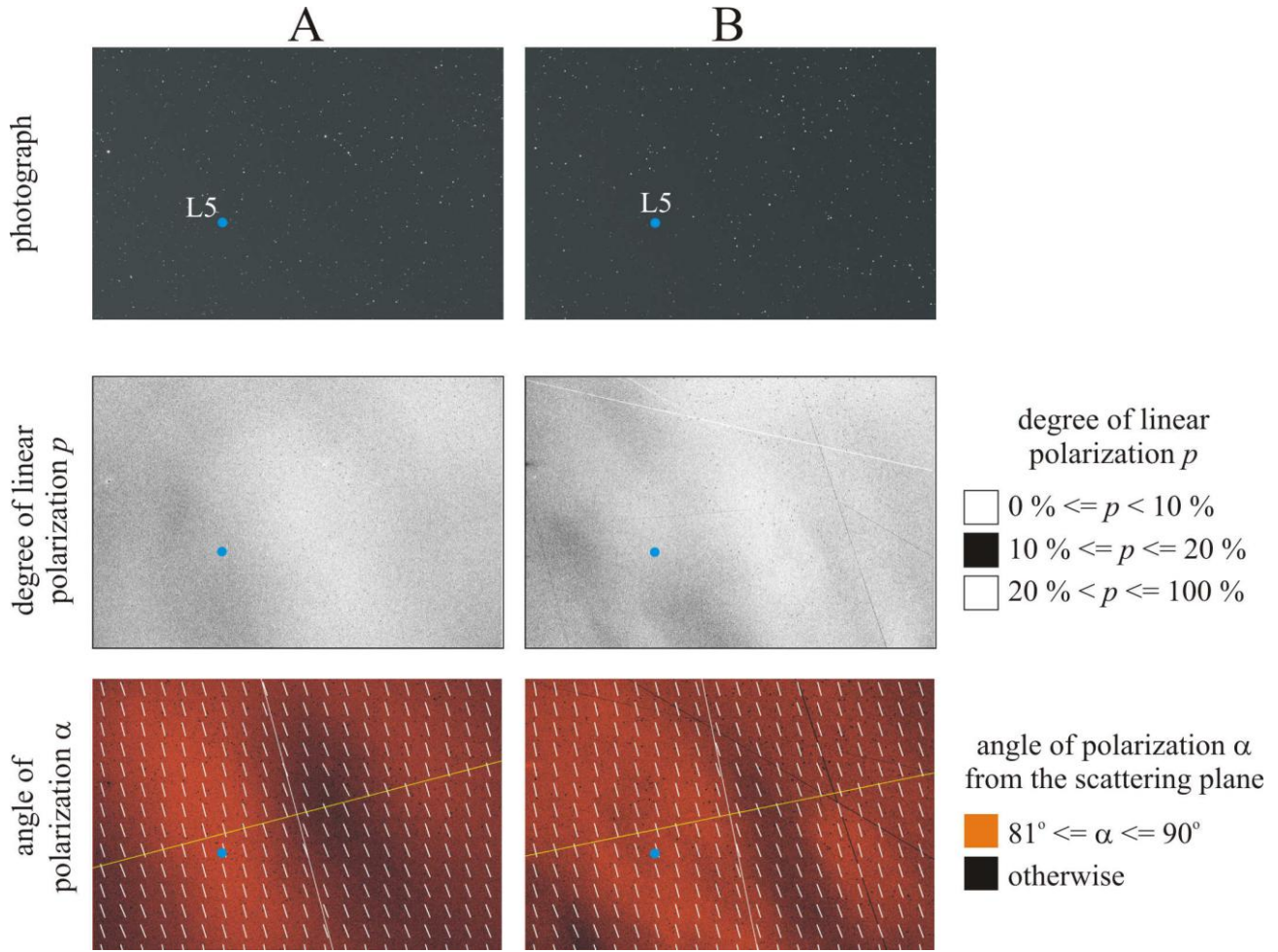

**Figure 2**: (A) Colour photograph, and patterns of radiance *I*, degree of linear polarization *p* and angle of polarization α (clockwise from the scattering plane) of the sky around the L5 Lagrange point of the Earth-Moon system measured by imaging polarimetry in the green (550 nm) spectral range at 23:29:67 UT on 17 August 2017 (picture center: RA = 2 h 12 m 28.2 s, DE = $8^o$ 3' 52.6") (A), and at 01:14:15 UT on 19 August 2017 (RA = 3 h 11 m 23.36 s, DE = $12^o$ 21' 15.38") (B). The position of the L5 point is shown by a blue dot. In the α-patterns the short white bars represent the local directions of polarization, while the long yellow and white straight lines show the scattering plane and the perpendicular plane passing through the center of the picture, respectively. The Kordylewski dust cloud is visible in both the *p*-pattern (clusters of black pixels with 10 % ≤ *p* ≤ 20 %) and the α-pattern (red pixels with $81^o$ ≤ α ≤ $90^o$). The *I*-, *p*- and α-patterns are very similar in the red (650 nm) and blue (450 nm) spectral ranges. Apart from the perpendicular white and yellow straight lines, the straight tilted lines in the *p*- and α-patterns of the B slides are traces of satellites.

Figure 3 shows the frequency of the degree of linear polarization *p* (%) measured around the L5 Lagrange point on 17 August 2017 at 23:29:67 UT with $73^o$ phase angle (Figure 3(A)), and on 19 August 2017 at 01:14:15 UT with $87.3^o$ phase angle (Figure 3(B)). The most frequently occurring *p*\*-value is 6 and 7 % in the former and latter case, respectively. Furthermore, the frequency of *p*\* in the former case is smaller (15.8 %) than that in the latter case (16.9 %).

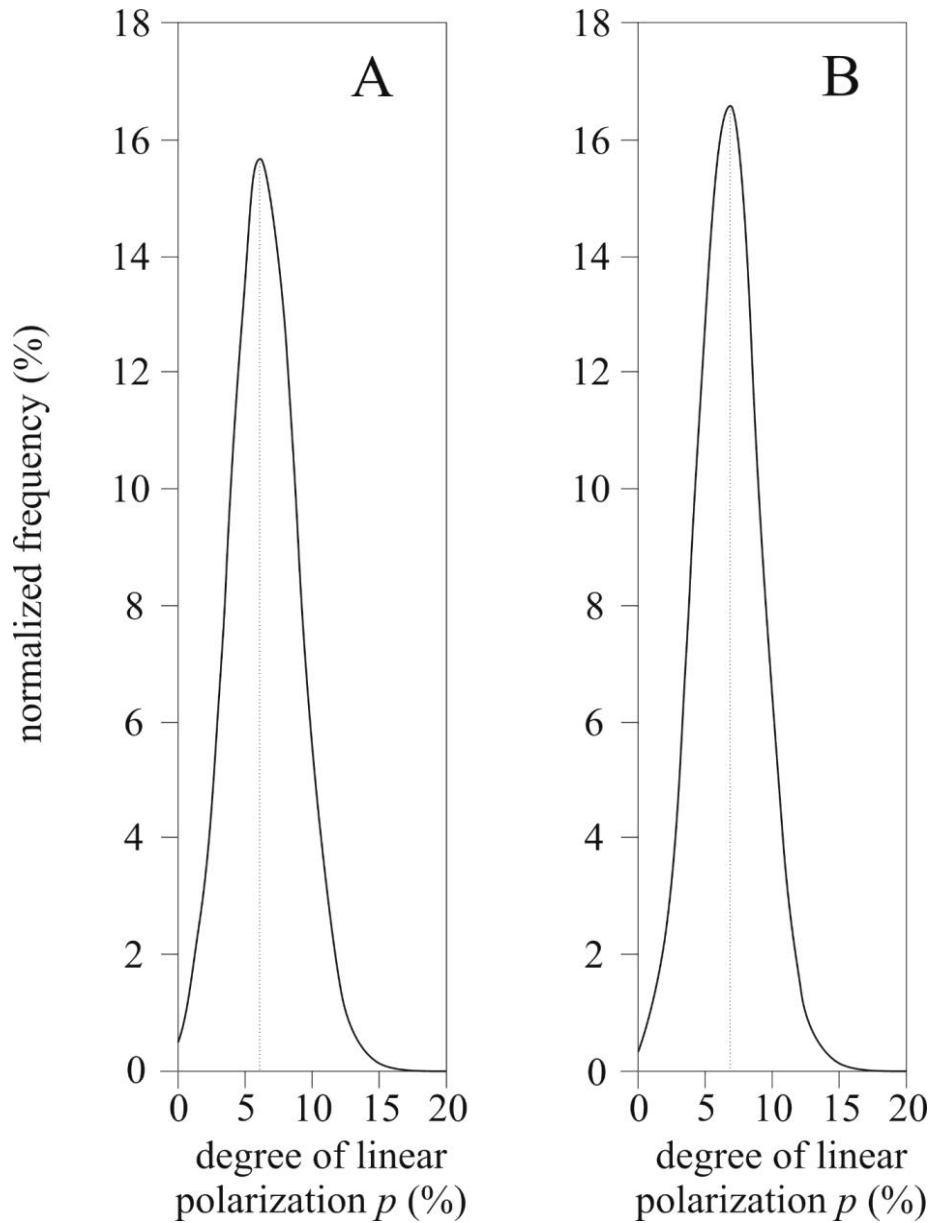

**Figure 3**: Normalized frequency (% = number of pixels with a given *p*-value divided by the total number of pixels of the pattern) of the degree of linear polarization *p* (%) measured around the L5 Lagrange point of the Earth-Moon system on 17 August 2017 at 23:29:67 UT with 73° phase angle (A), and on 19 August 2017 at 01:14:15 UT with 87.3° phase angle (B).

Figure 4(A) shows the *I*-, *p*- and α-patterns measured in the green (550 nm) at 23:43:17 UT on 16 October 2017 when the L5 point was not within the same celestial window as that of Figure 2. All three patterns are structureless: apart from the stars, galaxies and nebulae, the radiance is homogeneous dark, the degree of polarization is very low (< 10 %) and the direction of polarization is approximately vertical, being not perpendicular to the scattering plane. These features indicate the absence of the KDC in this celestial window.

Figure 4(B) displays the *I*-, *p*- and α-patterns measured at 21:51:30 UT on 8 June 2017 when a thin cirrus cloud covered the L5 point and its immediate vicinity. The *I*-pattern is diffuse striped, the *p*-values are approximately zero (< 5 %), and the nearly vertical directions of polarization are not perpendicular to the scattering plane. These optical characteristics do not correspond to those expected for a light-scattering particle cloud around the L5 point. Here the faint sunlight scattered by the KDC was overwhelmed by the faint ambient light scattered diffusely from the ice particles of the cirrus cloud.

Figure 4(C) represents the *I*-, *p*- and α-patterns measured at 20:59:58 UT on 15 July 2017 when condensation trails of an airplane occurred near the L5 point. Both in the colour photo and the *I*-pattern, the slowly dispersing light-scattering condensation trails drifting toward the top right corner of the picture/pattern are seen. In the *p*-pattern, the condensation trails occur as three distinct stripes due to its continuous drift. This is the result of the well-known motion artefact characteristic to sequential imaging polarimetry measuring moving/changing objects. Similar stripes occur in the α-pattern with inhomogeneous distribution of the direction of polarization being generally not perpendicular to the scattering plane. All these optical characteristics are quite different from those of the KDC observed by us (Figure 2).

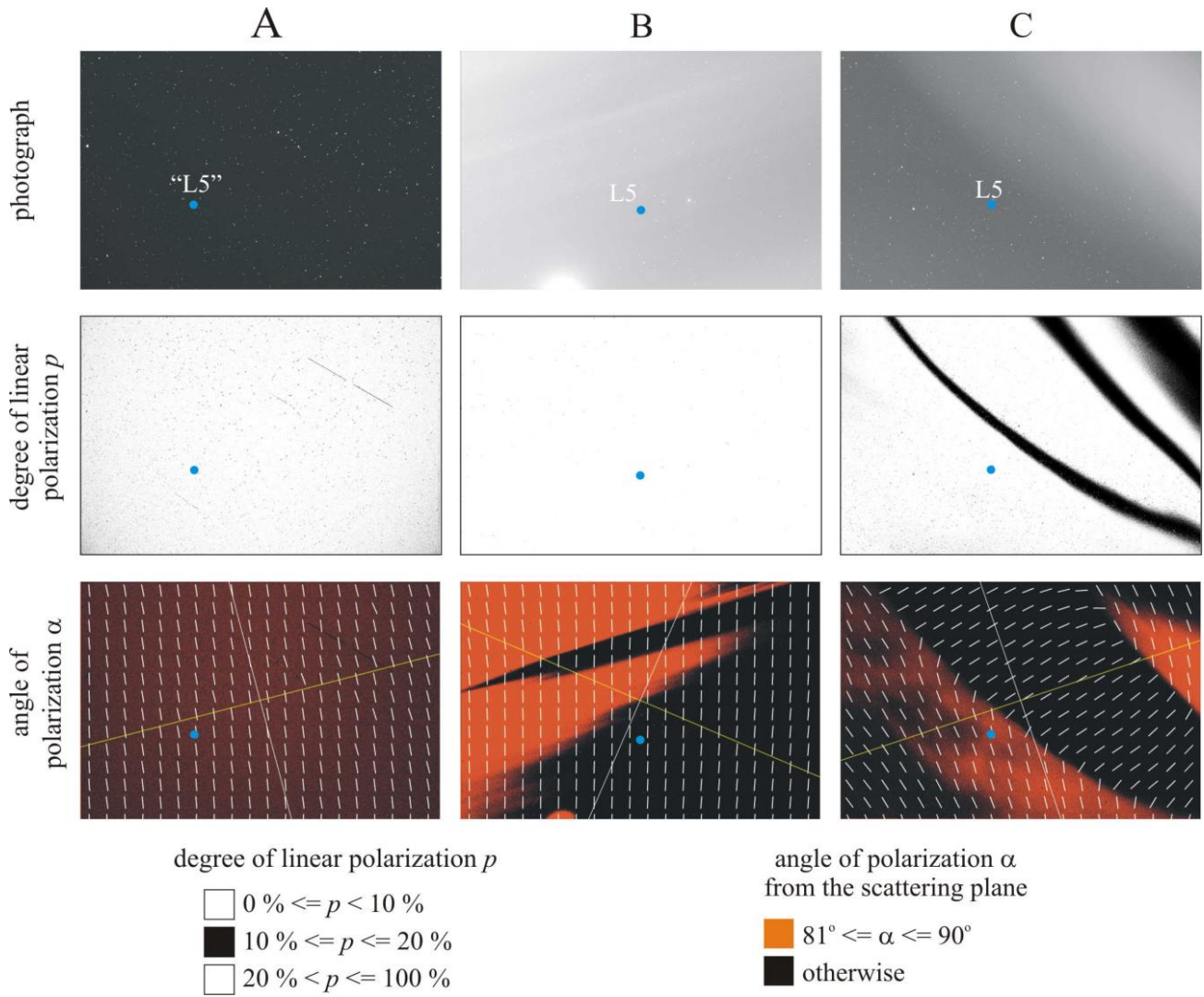

**Figure 4**: (A) As Figure 2 for the measurement performed at 23:43:17 UT on 16 October 2017 when the L5 point was not within this celestial window, therefore the blue dot "L5" shows the L5 position at 23:29:67 UT on 17 August 2017 (see Figure 2(A)). (B) Measurement at 21:51:30 UT on 8 June 2017 when a thin cirrus cloud covered this celestial window with the L5 point. (C) Measurement at 20:59:58 UT on 15 July 2017 when condensation trails of an airplane occurred in this celestial window with the L5 point. Apart from the perpendicular white and yellow straight lines, the straight tilted lines in the *p*- and α-patterns are traces of satellites.

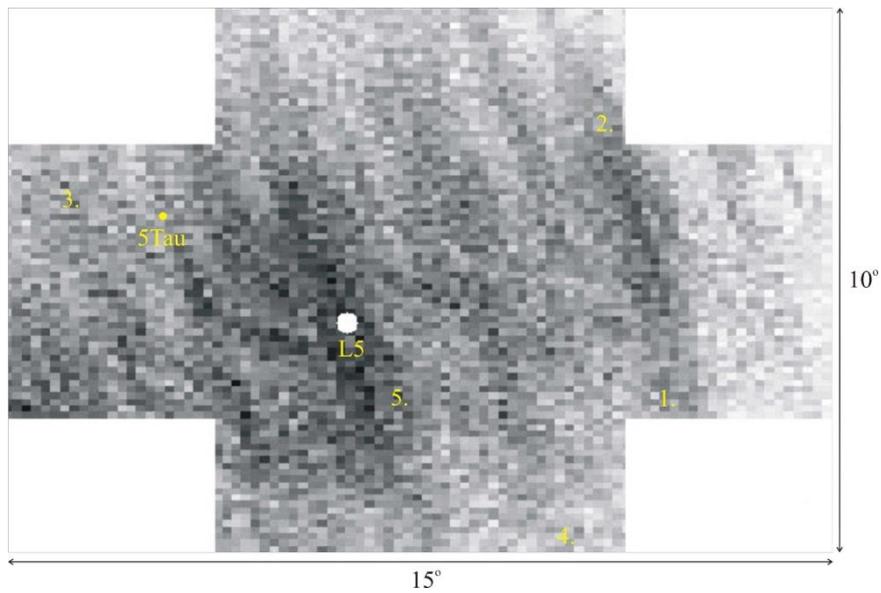

**Figure 5**: Computer-simulated volume density distribution of the particles of the KDC around the L5 point (white dot) of the Earth-Moon system. The darker the grey shade, the larger is the particle density. The numbered windows correspond to the fields of view of our imaging polarimetric telescope with which the polarization patterns of the sky around L5 were measured.

Figure 5 shows the computer-simulated volume density distribution of particles of the KDC around the L5 point, where the numbered windows correspond to the fields of view of our imaging polarimetric telescope with which the polarization patterns of the sky around L5 were measured.

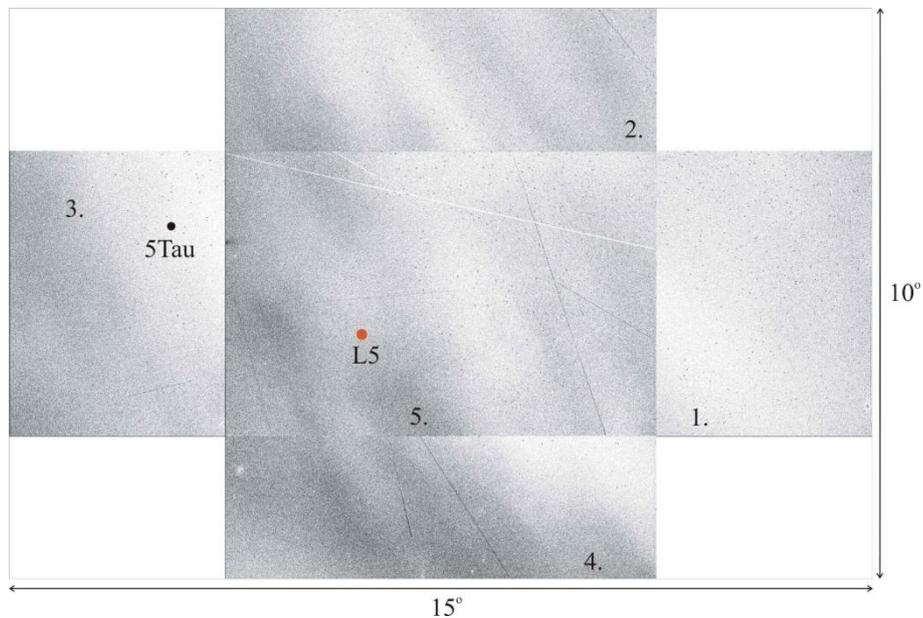

**Figure 6**: Mosaic pattern of the degree of linear polarization *p* of the KDC around the L5 point (red dot) measured on 19 August 2017 in the green (550 nm) with imaging polarimetry. The mean time (UT) of the patterns are: (1) 00:03:34, (2) 00:26:51, (3) 00:50:25, (4) 01:02:26, (5) 01:14:15.

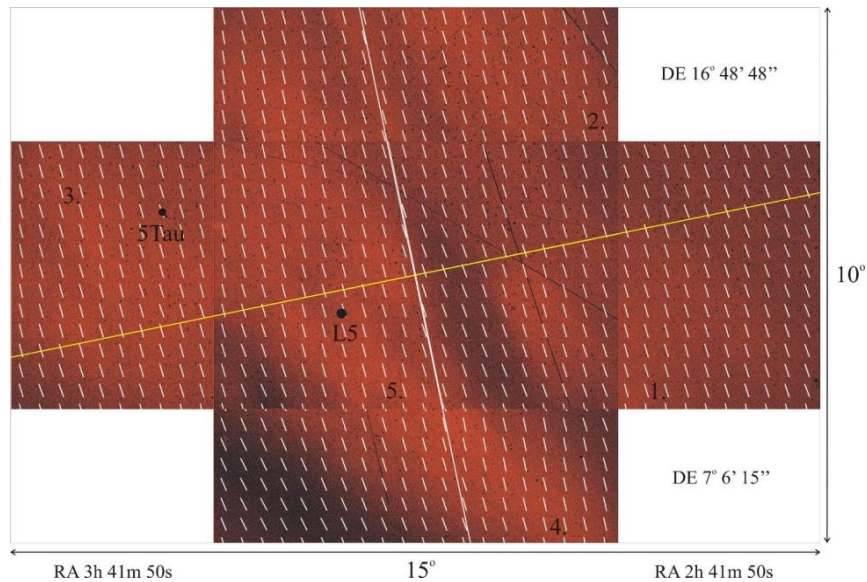

**Figure 7**: As Figure 5 for the angle of polarization α (clockwise from the scattering plane). The short white bars represent the local directions of polarization. The long yellow and white straight lines show the scattering plane and the perpendicular plane passing through the center of the picture, respectively.

Figures 6 and 7 display the mosaic patterns of the degree of polarization *p* and angle of polarization α of the KDC around L5 measured in the green (550 nm) with imaging polarimetry on 19 August 2017. Comparing the simulated particle density and the measured polarization patterns, a remarkable similarity can be seen: in all three patterns a multipartite structure occurs with several elongated clusters, showing that the KDC is a heterogenous particle cluster. The polarization patterns of the different neighbouring windows cannot be exactly fitted, because the sequential polarimetric measurements happened in slightly different points of time due to the necessary exposure (3×180 s), and during this short period the structure of the dynamic dust cloud slightly changed.

## 4. Discussion

Theoretically, there are extended small-concentration particle clouds around the L4 and L5 Lagrange points of the Earth-Moon system. Although the first mention of the possible accumulation of the zodiacal dust near the L2 point of the Sun-Earth system goes back to Moulton (1900), Kordylewski (1961) was the first to photograph two faint patches near the L5 point from the Polish mountain Kasprowy Wierch between 6 March and 6 April 1961. During his observation time, these patches with an angular diameter of about $6^o$ were slightly displaced relative to the L5 point. Since that time, these patches are believed by some scientists to be the KDCs. However, it is very difficult to detect the KDCs against the galactic light, star light, zodiacal light and skyglow (Roach 1975).

In spite of the pioneer observation by Kordylewski (1961) the existence of the KDCs is still under dispute, due to their extreme faintness making it difficult to confirm their existence. So far, there was no any convincing observational result, because the KDC is a very faint phenomenon, and it is also difficult to distinguish it from the even fainter zodiacal light. The latter is the sunlight scattered by the zodiacal dust. In the region of the antisolar point, the intensity of the zodiacal light is relatively enhanced, because each dust particle is seen in full phase. This phenomenon is the gegenschein (counterglow). So, it seems also the most convenient to photograph the KDC when it is near the antisolar point (full phase). However, in this case the polarization signature of the KDC is the weakest, consequently, its polarimetric study is the most difficult.

Over the past decades, some contradictory results have been achieved: Roosen (1966, 1968) found no evidence to the existence of KDCs near the L4 and L5 points. He suggested that if the KDCs exist at all, they are not associated with the Earth-Moon libration points. Wolff et al. (1967) did not find excess light in excess of 5 % of the light of the neighbouring night sky near the Lagrange points L4 and L5 of the Earth-Moon system, even though they photographed under astronomically favorable circumstances from an aircraft. However, Vanysek (1969) reported a successful visual observation (with naked eye of numerous persons) from an aircraft organized four times by NASA in 1966. The observers on that airplane described very faint nebulosities near the L4 and L5 points at large phase angles (at Vanysek (1969) the phase angle of the antisolar point is $180^o$). Vanysek (1969) proposed to detect the KDC during and shortly after the new-Moon phase, at small phase angles because of the strong forward scattering of sunlight by cloud particles. However, our polarization patterns suggest backward scattering, because the KDC in Figure 2(A) being nearer to the antisolar point is brighter than the KDC in Figure 2(B) being farther away. This may also indicate a silicate or limonite excess in the KDC over the ice or iron-like or slightly absorbing spherical particles (Vanysek 1969).

The KDC may be a transient phenomenon, because the L4 and L5 points might be unstable due to perturbations of the Sun, solar wind and other planets, as many astronomers believe. According to our computer simulations (Slíz-Balogh et al. 2018), the KDC has a continuously changing, pulsing and whirling shape, furthermore, the probability of dust particles being trapped is random due to the occasional incoming of particles and their incidental velocity vectors. Therefore, the structure and particle density of the KDC is not constant. The above-mentioned contradicting photometrical observations (Kordylewski 1961; Roosen 1966, 1968; Wolff et al. 1967; Vanysek 1969) also hint at the possible transient feature of the KDC.

Computer simulations investigated the problem of stability of the L4 and L5 points of the Earth-Moon system (Slíz et al. 2017; Salnikova et al. 2018; Slíz-Balogh et al. 2018). However, at a lunar eclipse the KDC could not be observed at all (Bruman 1969). A photographic search (Valdes & Freitas 1983) did not find any objects at the Earth-Moon Lagrange points L4 and L5. The limiting magnitude for the detection of libration objects near L4 and L5 was 17-19th magnitude. Thus, this survey was not sensitive enough to detect such diffuse clouds such as the KDCs. The Japanese Hiten space probe (using the Munich Dust Counter, an impact ionization detector designed to determine mass and velocity of cosmic dust) has passed through the L4 and L5 points of the Earth and Moon system, but did not find an obvious increase in dust concentration compared to the surrounding space (Igenbergs et al. 2012).

In spite of these negative results, there are, however, some positive reports about the photometric observations of the KDC. Analyzing the data from the Rutgers OSO-6 Zodiacal Light Analyzer experiment, Roach (1975) concluded that these dust clouds do exist in the L4 and L5 points, their angular size is about $6^o$ as seen from the Earth, and they move around the libration points. Using a number of parallel cameras at the observing station Roztoki Górne, Winiarski (1989) determined that the colours of the dust clouds near the L4 and L5 points differ from those of the counterglow (gegenschein), which means that the dust particles constituting them are also different.

According to our computer simulations (Slíz-Balogh et al. 2018), the KDC around the Lagrange point L5 of the Earth-Moon system is a dynamic structure with inhomogeneous, temporally changing particle density composed of several particle clusters. Since this dust cloud is illuminated by direct sunlight, the faint light scattered from the dust particles can be observed and photographed from the Earth surface with appropriately radiance-sensitive detectors. Such a pioneer photographical documentation has been first performed by Kordylewski (1961). According to the other above-mentioned successful trials (Vanysek 1969; Roach 1975; Winiarski 1989), the KDC can be visually detected only from small phase angles (determined by the observer, the Sun and the L4/L5 point), i.e. at or near "full dust moon". In this case the degree of polarization *p* of dust-scattered sunlight is minimal, practically zero. Since at phase angles near to $90^o$ the *p* of dust-scattered sunlight is maximal, it gives us the best chance to polarimetrically detect the KDC under this condition. Using imaging polarimetry, we indeed detected the polarization signature of the KDC in the L5 point of the Earth and Moon (Figure 2). Furthermore, the faint light scattered by the KDC can also be discerned in the colour photographs and the patterns of radiance *I* measured by us in the red, green and blue spectral ranges (Figure 2).

Theoretically, dust-scattered sunlight becomes partially linearly polarized with the direction of polarization perpendicular to the scattering plane determined by the Sun, the ground-born observer and the dust region observed (Können 1985; Coulson 1988; Collett 1994). We have indeed found this forecasted characteristic in the patterns of the angle of polarization measured with imaging polarimetry (Figures 2 and 7). This is one of the strongest proof that we observed a sunlit light-scattering object outside the Earth's atmosphere, rather than a terrestrial phenomenon. A further fact supporting the observation of the KDC is that in the measured α-patterns several clusters occur, as our simulations suggest (Slíz-Balogh et al. 2018).

Theoretically, the closer the angle of scattering is to $90^o$, the higher the degree of polarization *p* is of scattered light. We really found that the *p*-values of the KDC observed at 01:14:15 UT on 19 August 2017 with $87.3^o$ phase angle are higher than those observed at 23:29:67 UT on 17 August 2017 with $73^o$ phase angle (Figures 1(A), 1(B), 2(A), 2(B), 3). This is further convincing evidence that we registered the KDC with imaging polarimetry, rather than another phenomenon.

In order to exclude the possibility that with our polarization-sensitive telescope we registered an artificial optical phenomenon rather than the KDC we performed control measurements. We could imagine only the following three artifact possibilities:

- Unwanted ambient lights from the immediate optical environment reflected within our telescope from certain mechanical and/or optical elements resulted in the patterns which were interpreted as optical traces of the KDC. This possibility was excluded by our control measurement performed at 23:43:17 UT on 16 October 2017 when the L5 point was not within the celestial window studied, and indeed the typical, theoretically forecasted characteristics of the KDC were not registered (Figure 4(A)).
- A thin cloud covered the sky region studied, and its optical characteristics were assumed to origin from the KDC. This explanation was eliminated by the measurement done at 21:51:30 UT on 8 June 2017 when a cirrus cloud covered the investigated celestial window with the L5 point (Figure 4(B)): since the polarization characteristics of cirrus and other clouds are quite different from those expected for a KDC, the latter can easily be distinguished from the former.
- Condensation trails of an airplane occurred within the field of view of our telescope. This was excluded by our measurement at 20:59:58 UT on 15 July 2017 when such condensation trails crossed the studied celestial window with the L5 point (Figure 4(C)). Due to the typical striped polarization artefacts resulting from condensation trails, such trails can trivially be distinguished from the KDC.

If the KDC observed by us polarimetrically were part of the scattering of sunlight by the background zodiacal dust in the ecliptic (called zodiacal light), then the *p*- and α-patterns of Figure 4(A) should be quite similar to those of Figure 2. Since this is not true, this proves that the phenomenon is not the result of the background dust in the ecliptic.

The question may also arise whether the volume concentration of the KDC's particles is large enough to be detected on the ground. Due to the trapping effect of the Lagrange point L5, the particle density of the KDC should be greater than that of the surrounding zodiacal dust. If the latter has been photometrically observed from the ground, it means that the former is also detectable optically (photometrically and/or polarimetrically). The first evidence for the existence of the zodiacal dust and its band-structure was provided by the Infrared Astronomical Satellite (IRAS), while its first ground-based photometric observation was performed in 1997 by Ishiguro *et al.* (1999). They detected five very faint zodiacal bands, and emphasized that the ground-based photometry of the zodiacal light by a cooled CCD camera enabled them to investigate the structure and the temporary changes of these dust bands.

All ground-based observing systems are confronted with the light pollution of manmade ground-born light sources. These artificial lights usually increase the degree of linear polarization of skylight due to atmospheric aerosols (Kyba et al. 2011). Shkuratov et al. (2007), Kocifaj (2008) and Kocifaj et al. (2008) investigated the optical properties of these aerosol particles and their effect on light polarization. The photometric and polarimetric laboratory measurements of different surfaces and aerosol particles performed by Shkuratov et al. (2007) demonstrated the so-called negative polarization induced by the multiple scattering of light on rough surfaces and aerosols. Kocifaj et al. (2008) examined and compared the linear polarization of light scattered by homogeneous-sphere particles and Gaussian-core particles. Kocifaj (2008) carried out light pollution simulations and concluded that the role of ground-based light sources in light pollution is considerably enhanced under overcast sky conditions. The location of our imaging polarimetric measurements (Badacsonytördemic, $17°28'15"$ E, $46°48'27"$ N) is far away from all major settlements and there were only some local minor point sources (lamps), which were the same for all measurements, including the control measurement (without the L5 Lagrange point). Furthermore, during our measurements the sky was clear, cloudless. Thus, the effect of aerosol-induced light pollution on the measured polarization patterns was negligible during our measurements.

The direction of polarization of skyglow is perpendicular to the plane determined by the observer (polarimeter), the skyglowing celestial point observed and the ground-born light-polluting source (e.g. city lights). This direction is quite different from the measured direction of polarization of the KDC, which is perpendicular to the scattering plane (marked with a yellow straight line in Figure 2) determined by the observer (polarimeter), the Sun and the L5 Lagrange point. Due to the minimal light pollution in our measurement site, a relevant contribution of skyglow to the measured polarization signature was out of question. A minimal skyglow could have appeared only near the horizon, but the field of view of our imaging polarimetric telescope was far from the horizon. Thus, skyglow effects were surely negligible.

It is well known that multiple scattering of light can reduce the degree of polarization $p$ (Gehrels 1974; Coulson 1988; Horváth & Varjú 2004; Horváth 2014). Atmospheric turbidity, i.e. the spatio-temporal change of aerosol concentration can induce spatio-temporal changes of multiple scattering and thus those of $p$-values. Although a minimal atmospheric turbidity might have also occurred during our polarimetric measurements, to detect the existence of the polarization signature of the KDC and its non-existence in a control measurement (when the L5 Lagrange point was not within the field of view of our imaging polarimetric telescope), we selected only three nights (23:29:67 UT on 17 August 2017 - Figure 2(A), 01:14:15 UT on 19 August 2017 - Figure 2(B), 23:43:17 UT on 16 October 2017 - Figure 4(A)) when the sky was clear, without a visually detectable atmospheric turbidity. Such turbidities occurring under clear sky conditions could result in only slight (< 5 %) changes in the degree of polarization $p$ (Horváth & Varjú 2004, Horváth 2014). However, the maximum of the registered $p$-values in Figures 2 and 3 is about 20 %. Such large $p$-variations cannot be induced by the aerial turbidity under clear skies. Consequently, the significant differences between the polarization patterns of Figure 2 (KDC) and Fig 4(A) (control) cannot be the result of atmospheric turbidities.

On the basis of the above arguments we conclude that for the first time we have observed and registered polarimetrically the Kordylewski dust cloud around the Lagrange point L5 of the Earth and Moon. By this we corroborated the existence of the KDC first observed photometrically by Kordylewski (1961).

Similarly, to many objects and optical phenomena in nature, the knowledge of polarization characteristics can provide valuable additional information. Although the KDC can also be observed with radiance-sensitive devices, it can be registered easier and more effectively and studied by polarization-sensitive telescopes like our one. The observability of the KDC is different from that of the Arago, Babinet, Brewster and fourth neutral points of the atmosphere: these unpolarized celestial points cannot be seen at all in colour photographs or radiance patterns measured in different spectral ranges, but can be observed and studied in the patterns of the degree and angle of polarization of skylight (Horváth et al. 2002).

In the future, it would also be worth studying both computationally and imaging polarimetrically the dynamical and optical characteristics of the KDC around the L4 Lagrange point of the Earth-Moon system. It would be interesting to compare the features of the KDCs formed around the L5 and L4 points. For these tasks several polarization-blind astronomical telescopes should be mounted with imaging polarimetric devices composed of rotatable linear polarizers. One could also try to measure the circular polarization (if any) of the light scattered by the KDCs with an appropriate polarimeter.

The existence of the KDC suggests the challenging possibility that appropriate astronautical missions could take samples from the particles librating at and around the L4 and L5 points of the Earth and Moon. The investigation of these clouds could be important from the point of view of space navigation safety.

**Author Contributions**

Substantial contributions to concept and design: JSB, AB, GH
Performing the experiments and data acquisition: JSB, AB
Data analysis and interpretation: JSB, GH
Drafting the article and revising it critically for important intellectual content: JSB, GH

**Acknowledgements**

We are grateful to Attila Mádai (mechanical engineer, Érd, Hungary) for his valuable technical help. We thank Tamás Hajas and Edward Brown for improving the English of our paper. We thank an anonymous reviewer and both the scientific and assistant editors for their constructive and valuable comments on earlier versions of our manuscript.

**Funding**

No funding is declared.

## References

Arago, D. F. J. 1811, Mém. Cl. Sci. Math. Phys. 1: 93-134

Azzam, R. M. A., Bashara, N. M. 1992, Ellipsometry and Polarized Light. North-Holland, Amsterdam, New York

Babinet, J. 1840, Comptes Rendus 11: 618-620

Belskaya, I. N., Bagnulo, S., Stinson, A., et al. 2012, A&A 547: id.A101 (doi: 10.1051/0004-6361/201220202)

Brewster, D. 1842, Rep. Brit. Assoc. Adv. Sci. 2: 13-25

Brewster, D. 1847, Phil. Magaz. J. Sci. 31: 444-454

Bruman, J. R. 1969, Icarus 10: 197

Collett, E. 1994, Polarized Light: Fundamentals and Applications. Marcel Dekker Inc, New York

Coulson, K. L. 1988, Polarization and Intensity of Light in the Atmosphere. A. Deepak Publishing, Hampton, Virginia, USA

Danby, J. M. A. 1992, Fundamentals of Celestial Mechanics. 2nd ed. Richmond: Willman-Bell

Gehrels, T. (ed.) 1974, Planets, Stars and Nebulae Studied with Photopolarimetry. Univ Arizona Press, Tucson, Arizona

Hadamcik, E., Sen, A. K., Levasseur-Regourd, A. C., et al. 2014, Meteoritics and Planetary Science 49: 36

Horváth, G., Bernáth, B., Suhai, B., et al. 2002, J. Opt. Soc. Am. A 19: 2085-2099

Horváth, G. (ed.) 2014, Polarized Light and Polarization Vision in Animal Sciences. Springer: Heidelberg, Berlin, New York

Horváth, G., Varjú, D. 2004, Polarized Light in Animal Vision - Polarization Patterns in Nature. Springer: Heidelberg, Berlin, New York

Igenbergs, E., Hüdepohl, A., Uesugi, K., et al. 2012, Chapter 9: The Munich dust counter - A cosmic dust experiment on board of the Muses-A mission of Japan. pp. 45-48. In: Origin and Evolution of Interplanetary Dust. (editors: A. C. Levasseur-Regourd, H. Hasegawa) Kluwer Academic Publishers

Ishiguro, M., Nakamura, R., Fujii, Y., et al. 1999, ApJ 511: 432

Ivanova, O., Shubina, O., Moiseev, A., et al. 2015, AstBu 70: 349-354

Kocifaj M. 2008, ApOpt 47: 792

Kocifaj, M., Kundracik, F., Videen, G. 2008, J Quant Spect Rad Trans 109: 2108-2123

Kyba, C. C. M., Ruhtz, T., Fischer, J., et al. 2011, J Geophys Res D (Atmospheres) 116 (D24): D24106 (doi: 10.1029/2011JD016698)

Kordylewski, K. 1961, actaa 11: 165

Können, G. P. 1985, Ciel et Terre 102: 63

Marin, F., Goosmann, R. W., Dovciak, M., et al. 2012, MNRAS 426: L101-L105 (doi: 10.1111/j.1745-3933.2012.01335.x)

Marin, F., Muleri, F., Soffitta, P., et al. 2015, A&A 576: id.A19 (doi: 10.1051/0004-6361/201425341)

Moulton, F. R. 1900, AJ 21: 17-22

Reig, P., Blinov, D., Papadakis, I., et al. 2014, MNRAS 445: 4235-4240 (doi: 10.1093/mnras/stu2322)

Roach, J., 1975, P&SS 23: 173

Roosen, R. G. 1966, Sky & Telescope 32: 139

Roosen, R. G. 1968, Icarus 9: 429-439

Roosen, R. G., Harrington R. S., Jeffreys W. H. 1967, PhT 20: 9-11

Salnikova, T., Stepanov, S., Shuvalova, A. 2018, ActaAstro, in press (doi: 10.1016/j.actaastro.2017.12.022)

Shkuratov Y., Bondarenko S., Kaydash V., et al. 2007, JQSRT 106: 487

Simpson, J. W. 1967, PhT 20: 39-46

Slíz, J., Kovács, T., Süli, Á. 2017, Astron Nach 338: 536

Slíz-Balogh, J., Barta, A., Horváth, G. 2018, MNRAS 480: 5550-5559 (doi: 10.1093/mnras/sty2049)

Steg, L., De Vries, J. P. 1966, SSR 5: 210-233

Szebehely, V. 1967, New York: Academic Press, 1967

Valdes F., Freitas R. A. Jr. 1983, Icarus 53: 453-457

Vanysek V. 1969, Nature 221: 47

Winiarski M. 1989, EM&P 47: 193-215 (doi: 10.1007/BF00058142)

Wolff C., Dunkelman L., Hanghney L. C. 1967, Sci 157: 427-429

Zejmo M., Slowikowska A., Krzeszowski K. et al. 2017, MNRAS 464: 1294-1305 (doi: 10.1093/mnras/stw2335)